\documentclass[11pt]{article}
\usepackage{graphicx,wrapfig,color}
\usepackage{subfig}
\usepackage{setspace}
\usepackage{epsfig}
\usepackage{graphicx}
\usepackage{multicol}
\usepackage{amsfonts}
\usepackage{latexsym}
\usepackage{amsmath,amssymb}
\usepackage{enumerate}
\usepackage{cite}
\usepackage{lipsum}
\usepackage{mwe}
\usepackage{subfig}
\newcommand{\ben}{\begin{enumerate}}
\newcommand{\een}{\end{enumerate}}
\newcommand{\spa}{\phantom{s}}
\newcommand{\bea}{\begin{eqnarray}}
\newcommand{\eea}{\end{eqnarray}}
\newcommand{\be}{\begin{equation}}
\def\bel#1{\begin{equation} \label{#1}}
\newcommand{\ee}{\end{equation}}
\newcommand{\bi}{\begin{itemize}}
\newcommand{\ei}{\end{itemize}}
\newcommand{\ba}{\begin{align}}
\newcommand{\ea}{\end{align}}

\def\bel#1{\begin{equation} \label{#1}}

\def\b{\beta}

\def\mc{\mathcal}
\def\be{\begin{equation}}
\def\ee{\end{equation}}
\def\bea{\begin{eqnarray}}
\def\eea{\end{eqnarray}}

\def\ltap{\ \raise.3ex\hbox{$<$\kern-.75em\lower1ex\hbox{$\sim$}}\ }
\def\gtap{\ \raise.3ex\hbox{$>$\kern-.75em\lower1ex\hbox{$\sim$}}\ }
\def\gl{\ \raise.5ex\hbox{$>$}\kern-.8em\lower.5ex\hbox{$<$}\ }
\def\roughly#1{\raise.3ex\hbox{$#1$\kern-.75em\lower1ex\hbox{$\sim$}}}

\def\nn{\nonumber}

\def\spa{\phantom{a}}

\def\cv{{{\cal{V}}}}

\newcommand{\comments}[1]{}
\usepackage{color}
\definecolor{cblue}{RGB}{100,5,255}
\definecolor{cred}{RGB}{255,50,40} 
\definecolor{cgreen}{RGB}{40,255,40} 
\definecolor{corange}{RGB}{250,200,40}

\begin{document}

\begin{titlepage}
\vskip 1 cm
\begin{center}
{\Large \bf  Constraints on K\"ahler moduli inflation from reheating
}
\vskip 1.5cm  
{ 
Sukannya Bhattacharya$^{*}$, Koushik Dutta$^{*}$, Anshuman Maharana$^{\dagger}$
\let\thefootnote\relax\footnotetext{ \hspace{-0.5cm} E-mail: {$\mathtt{sukannya.bhattacharya@saha.ac.in, koushik.dutta@saha.ac.in, anshumanmaharana@hri.res.in} $}}
}
\vskip 0.9 cm

{\textsl{$^{*}$Theory Division, \\
 Saha Institute of Nuclear Physics, \\
HBNI, 1/AF Salt Lake,  \\
 Kolkata - 700064, India.}
}
\vskip 0.6 cm
{\textsl{
$^{\dagger}$Harish Chandra Research Intitute, \\
HBNI, Chattnag Road, Jhunsi,\\
Allahabad -  211019, India.\\}
}
%
%
\end{center}

\vskip 0.6cm

\begin{abstract}
\vskip 0.5 cm
{We present  predictions of the  K\"ahler moduli inflation model for the spectral tilt by parametrising the reheating epoch by an effective equation-of-state parameter and the number of e-foldings of reheating; and taking into account the post-inflationary history of the model. This model has an epoch in which the energy density of the universe is dominated by cold moduli particles. We compare our results with data from the PLANCK mission and find that exotic reheating (with effective equation of state $w_{\rm re}$ greater than 1/3) or dark radiation is required to match the observations. For canonical reheating case with $w_{\rm re} = 0$, we deduce $\text{log}_{10}(T_{\rm re}/10^3~ \text{GeV}) \simeq 1190 (n_s - 0.956)$. We also analyse our results in the context of  observations being planned for the future and their projected sensitivities.}
\noindent

\end{abstract}

\vspace{3.0cm}

\end{titlepage}
\pagestyle{plain}
\setcounter{page}{1}
\newcounter{bean}
\baselineskip18pt
%



\section{\Large{Introduction}}
\label{Seckmi}

The observation of an adiabatic and almost scale invariant spectrum of inhomogeneities in the Cosmic Microwave 
Background (CMB) has given strong evidence in favour of the inflationary paradigm \cite{inflation1, planck2015}. In the next decade, it is expected
that more precision data will come in. Various observational programs plan to measure the scalar spectral tilt $n_s$ with improved accuracy and minutely probe the CMB for non-Gaussianities and tensor modes. For example, it is expected that the spectral index is going to be measured with an accuracy of $\Delta n_s \sim 0.002$ ($1$-$\sigma$) by forthcoming ground based CMB-S4 experiment \cite{Abazajian:2016yjj}, and satellite based experiment CORE \cite{Finelli:2016cyd}. If approved, it is expected that these experiments are going to be operational within ten years.  On the theoretical front, ultraviolet sensitivity of the slow roll parameters necessitates the embedding of inflationary models in a ultraviolet complete setting.  Thus it is natural to carry out inflationary model building in String Theory and  with the advent of precision data, it will be crucial to develop the necessary tools so that accurate  theoretical predictions can be made for these models.  

  The central input for obtaining  predictions of an inflationary model is the number of e-foldings between horizon exit and the end of inflation $(N_{*})$. This in turn depends  on the entire post-inflationary history of the 
universe (including reheating).  A generic feature of the post inflationary history of models of inflation constructed in string theory (and supergravity) is an epoch in which the energy density of the universe is dominated by cold moduli particles (which arises as a result of vacuum misalignment during the inflationary epoch) - See \cite{Kane:2015jia} for a recent review. Thus, in order to obtain precise theoretical predictions it is necessary  to incorporate the effect  of this epoch along with the reheating epoch. In this paper, we will carry out this analysis for   K\"ahler moduli inflation \cite{Kahler}.  K\"ahler moduli inflation is a model of inflation in the Large Volume Scenario  (LVS) for moduli stabilisation \cite{LVS, LVS2} in IIB flux compactifications \cite{GKP}. Recently, a `global' embedding of the model in compact oreintifold was provided in \cite{micu}. The post-inflationary history of this model was analysed  in \cite{BLH} --  in particular the dynamics of the  epoch in which the energy density is dominated by cold moduli particles was studied in detail; the effect of this epoch on the value of $N_{*}$ was computed.  Although in general the precise microscopic details of reheating can be complicated, we will follow the usual approach of parametrising the effect of the reheating epoch on $N_{*}$ by the number of e-foldings during reheating $(N_{\rm re})$ and the effective equation of state $(w_{\rm re})$ during the epoch (see for e.g. \cite{Dai:2014jja}). With this the inflationary predictions can be expressed in terms of $N_{\rm re}$ and $w_{\rm re}$. Interesting constraints arise from the fact  that $w_{\rm re}$ can not be arbitrary; physical arguments and simulations constrain the range of $w_{\rm re}$. Overall our analysis is similar in spirit to \cite{initial_approach},  subsequent analysis along these lines using PLANCK data  has been carried out in in \cite{Dai:2014jja, later_approach }. Recently, analysis similar to ours has been carried out for the fibre inflation model in \cite{lat}. We note that  fibre inflation does not lead to an epoch of modulus domination in the post inflationary history. In the case of K\"ahler moduli inflation this epoch plays a crucial role.

  This paper is organised as follows. We first review some basic aspects of K\"ahler moduli inflation and briefly outline the post-inflationary history of the model. Then, following \cite{BLH} we obtain the dependence of $N_{*}$ on the reheating parameters. We then obtain the expression for the spectral tilt $n_s$ in terms of the reheating parameters and compare the model predictions to observational data. Our results are summarised in 
the plots in section \ref{results}.  Given the number of e-foldings during the reheating epoch, the temperature at the end of reheating $T_{\rm re}$ of the Standard Model is determined; thus predictions for the spectral tilt in terms of $N_{\rm re}$ and $w_{\rm re}$ can be parametrised in terms of $T_{\rm re}$ and $w_{\rm re}$. As in \cite{Dai:2014jja}, we also present our results in terms of the later parametrisation.

  
\section{Review of K\"ahler Moduli Inflation} 
\label{Seclvs}


In this section, we briefly review K\"ahler moduli inflation and refer the reader to \cite{Kahler} for details of the model. As discussed in the introduction, the K\"ahler moduli inflation model is set in the  Large Volume Scenario (LVS) for moduli stabilisation \cite{LVS,LVS2}. We begin by  briefly reviewing LVS, which is set in the IIB flux compactifications and the complex structure moduli are stabilised by three form fluxes. The complex structure moduli can be integrated out, and the super potential for the K\"ahler moduli is given by 
\bel{superi}
 W = W_0 + \sum_{i} A_i e^{-a_i T_i}~,
\ee
where $W_0$ is the expectation value of the Gukov-Vafa-Witten super  potential \cite{gukov}, and the sum is over the K\"ahler moduli $T_i = \tau_i +  i c_i$ ($\tau_i$ are the volume of the four cycles and $c_i$ are the associated axionic partners). The K\"ahler moduli can then be stabilised following  the LVS algorithm for Calabi-Yaus with negative Euler number when  
$W_0$ is of order unity.  The key ingredient is  incorporating  the leading $\alpha'$ correction to the K\"ahler
potential given by \cite{BBHL}: $K = - 2 \ln \left( \cv + {\hat\xi \over 2} \right)$,
where $\hat\xi$ is a function  of the Euler number of the Calabi-Yau and the dilaton  vacuum expectation value  $\hat\xi =  \chi/(2 (2 \pi)^3 g_s^{3/2})$ (see \cite{RAF} for a derivation of the K\"ahler potential in the presence of seven branes). 

The simplest models of LVS (which will also be relevant for us) are the
ones in which the volume of the Calabi-Yau takes  the Swiss-cheese form: $\cv = \alpha \bigg( \tau_1^{3/2} - \sum_{i=2}^{n} \lambda_i \tau_i^{3/2} \bigg)$  \cite{LVS,LVS2}. For such models, the overall volume is controlled by $\tau_1$, the  moduli $\tau_2 ,..., \tau_n$ are blow-up modes, and their magnitudes give the size of the holes in the compactification. The scalar potential in the limit
$\cv \gg 1$ and  $\tau_1 \gg \tau_i \spa ( {\rm{for}} \spa  i >1)$ computed from the above superpotential  and K\"ahler potential is given by: 
\bel{total}
V_{\rm LVS} = \sum_{i=2}^n { 8 (a_i A_i)^2 \sqrt{\tau_i} \over 3 \cv \lambda_i } e^{-2 a_i \tau_i}
-  \sum_{i=2}^{n} { 4 a_i A_i W_0 \over \cv^2} \tau_i e^{-a_i \tau_i} + {3 \hat\xi  W_0^2 \over 4 \cv^3 } + { D  \over \cv^{\gamma} }\,.
\ee 
where the phases of the axions $c_i$ have been chosen so as to minimise the potential, and the last term $V_{\rm up} = { D  \over \cv^{\gamma} }$ with $D > 0$ and  $1\leq\gamma\leq 3$ has been added explicitly. Without $V_{\rm up}$, the potential 
has a minimum in the `large volume limit': $\cv \to \infty$ with  $a_i \tau_i \approx \ln \cv\,$ with a negative value of the vacuum energy. To have a solution with a vanishing  (or slightly positive) cosmological constant an `uplift term' $V_{\rm up}$ has been incorporated in the effective action (such a term can arise from various mechanisms for e.g. anti-D3 branes in warped throats \cite{kklt},  dilaton-dependent non-perturbative effects \cite{dil}, magnetised D7-branes \cite{mag1, mag2}, or the effect of D-terms \cite{rum}).  

The role of the inflaton is played by one of the blow-up moduli $\tau_n$. As mentioned before, we will focus on the case when the volume of the Calabi-Yau is of Swiss-Cheese type\footnote{The same analysis can be carried out for more general Calabi-Yaus following \cite{CD}. }.
The $\tau_n$ modulus is displaced from its global minimum during the inflationary epoch. In the regime $e^{a_n \tau_n} \gg \cv^2$ the potential of Eq.~\eqref{total}  can be approximated by: 
\bel{vap}
V_{\rm inf} = \sum_{i=2}^{n-1} { 8 (a_i A_i)^2 \sqrt{\tau_i} \over 3 \cv \lambda_i } e^{-2 a_i \tau_i}
-  \sum_{i=2}^{n-1} { 4 a_i A_i W_0 \over \cv^2} \tau_i e^{-a_i \tau_i}
+ {3 \hat\xi  W_0^2 \over 4 \cv^3 } + { D  \over \cv^{\gamma} } - { 4 a_n A_n W_0 \over \cv^2} \tau_n e^{-a_n \tau_n}\,. \nn
\ee
The inflaton ($\tau_n$) has an exponentially flat potential; and the other directions ($\cv, \tau_i$ with $i = 2, .. ,n-1$) in field space are heavy during inflation. Integrating out the heavy  directions, inflaton potential in terms of the canonical field $\sigma$ is 
\bel{canpot}
V = V_0 - \frac{4 W_0 a_n A_n}{\mc{V}^2_{\rm in}} \left(\frac{3 \mc{V}_{\rm in} }{4 \lambda} \right)^{2/3}  \sigma^{4/3}
\exp \left[-a_n \left(\frac{3 \mc{V}_{\rm in}}{4 \lambda}\right)^{2/3} \sigma^{4/3}\right]~,
\ee
where \bel{can}
\frac{\sigma}{M_{\rm pl}} = \sqrt{\frac{4 \lambda}{3 \mc{V}_{\rm in}}} \,\tau_n^{\frac{3}{4}} ~~~~{\rm with} \spa \spa V_{0} = \frac{\b W_0^2}{\mc{V}_{\rm in}^3}~,
\ee
and  $\cv_{\rm in}$ is the value of the volume modulus during the inflationary epoch and $\beta$ an $\mc{O}(1)$ constant. 

Phenomenological considerations including the constraints from the  strength of the amplitude of scalar perturbations require $\cv_{\rm in} \approx 10^{5}$ to $10^{6}$ \cite{BLH}. In this region of the parameter space the spectral tilt $(n_s)$ can be expressed in terms of the number of e-foldings between horizon exit and the end of inflation by the formula
\bel{nx}
  n_s \approx 1 - { 2 \over N_{*} }.
\ee
The tensor to scalar ratio is rather insensitive to $N_{*}$; $r \approx 10^{-10}$ to $10^{-11} $ is  in the phenomenologically viable range. Of course the  above expression of Eq.~\eqref{nx} for the spectral index is not exact, and in principle can be evaluated  by solving for the evolution of the inflation field numerically. The tensor to scalar ratio, $r$ also has mild dependence on the model parameters. Therefore theoretical predictions are sensitive to global embedding of the model in a compactification, and given a global embedding, numerical evolution of the fields has to be performed to obtain the predictions as in \cite{micu}.  For the present analysis we will take a phenomenological approach (as in \cite{BLH}) -- we will use the expression of Eq.~\eqref{nx} for $n_s$ whereas $r$  will be taken to be in the above range. Finally, we note  that the above expression for the spectral index and the value of tensor-to-scalar ratio are also essentially independent of the post-inflationary history of the universe.

  
\section{Post-inflationary History and Reheating}

  As mentioned in the introduction, the key feature of the post-inflationary history of the model that is relevant for our analysis is the epoch in which the energy
density is dominated by cold moduli particles. This arises as a result of vacuum misalignment; the volume modulus is displaced from its post-inflationary minimum during the inflationary epoch. This displacement was computed explicitly in
\cite{BLH} by analysing the scalar potential in the inflationary epoch. The displacement of the canonically normalised field in Planck units was found be $\mathcal{O}(0.1 M_{\rm pl}) $, in keeping with generic expectations from effective field theory. At the end of inflation, with the expansion of the universe  the Hubble friction term can no longer keep the volume modulus away from its post-inflationary minimum; the volume modulus begins to perform coherent oscillations about its post-inflationary minimum. The energy density associated with this falls off as $a^{-3}(t)$, thus it quickly dominates over the energy density associated with radiation produced from the decay products of the inflaton which falls off as $a^{-4}(t)$. This epoch of modulus domination lasts until the decay of the moduli particles.

   Now, let us come to the determination of $N_{*}$ for the model. In any cosmological model $N_{*}$ is determined by tracking the evolution of the energy density of the universe from the point of horizon exit of the CMB modes to the present epoch. The formula for the strength of density perturbations
$$
  A_s = { 2 \over 3 \pi^2 r } \bigg({ \rho_{*} \over M_{\rm pl}^{4} } \bigg)
$$
gives the energy  density of the universe at the time of horizon exit $(\rho_{*})$;  demanding that this energy density
evolves to the energy density observed today gives the equation that determines $N_{*}$. For the standard cosmological timeline (consisting of inflation, reheating, epoch of radiation domination, epoch of matter domination and finally the present epoch of dark energy domination) this yields
$$
    N_{*} + {1 \over 4} ( 1  - 3 w_{\rm re}) N_{\rm re} \approx 57 + { 1 \over 4} \ln r +{1 \over 4} \ln \bigg( {\rho_{*} \over \rho_{\rm end}} \bigg)~.
$$
For K\"ahler moduli inflation (or any cosmological model with a non-standard post-inflationary history), the equation determining $N_{*}$ gets modified. The equation determining $N_{*}$ for K\"ahler moduli inflation was obtained in \cite{BLH} (using the analysis of \cite{D})\footnote{In addition to the terms in Eq.~\eqref{one}, reference \cite{BLH} found a term associated with the number of e-foldings
in which cold inflaton particles dominate the energy density.  This term was found to be small in comparison with the other;  we will drop it in our analysis.}
\bel{one}
  N_{*} +  {1 \over 4} N_{\rm mod2} + {1 \over 4} (1 - 3w_{\rm re})N_{\rm re} \approx 57 + {1 \over 4} \ln r + {1 \over 4} \ln \bigg( {\rho_{*} \over \rho_{\rm end}} \bigg) ~.
\ee
Here $N_{\rm mod2}$ is the number of e-foldings that the universe undergoes during the epoch in which the energy density is dominated by the volume modulus. The  R.H.S of the above equation is entirely determined by the details of inflation. We note that the dependence on $r$ is mild, but the size of the term involving $r$ is appreciable as  K\"ahler moduli inflation has a very small value of $r$ $(r \approx 10^{-10})$.   Also, since the potential for K\"ahler Moduli inflation is exponentially flat, it is a good approximation to take ${\rho_{*} \over \rho_{end}} \approx 1$. Our ignorance about the detailed mechanism of the reheating epoch is parametrised by the effective equation of state parameter $w_{\rm re}$, and the number of e-folding during the epoch $N_{re}$. Of course, for a model  in which all the couplings between the modulus field with the Standard Model degrees of freedom are known, the mechanism for reheating can be determined and  $N_{\rm re}$ and $w_{\rm re}$ can be computed. In \cite{BLH}, post inflationary dynamics of the volume modulus was discussed in detail, and it was found that $N_{\rm mod2} \approx 25$. Taking these inputs, Eq.~\eqref{one} becomes
\bel{two}
 \frac{2}{1 - n_s} +  {1 \over 4} (1 - 3w_{\rm re})N_{\rm re} \approx 45~.
\ee
 The above equation will be central for our analysis to confront the model with the data in the next section. Before proceeding to this analysis, let us discuss some points which will play an important role.  
 
   Firstly, the range of the equation of state parameter $w_{\rm re}$. The simplest model for reheating is  the canonical reheating scenario -- the scalar field oscillates coherently around a quadratic minimum producing a cold gas of  particles, these  decay to the Standard Model sector producing a thermal bath of temperature $T_{\rm re} \sim \sqrt{\Gamma M_{\rm pl}}$ (where $\Gamma$ is the total decay width). This has $w_{\rm re} =0$. More generally, if the oscillations take place around a minimum of the form $\phi^n$ (with $n$ even), the equation of state parameter is given by $w_{\rm re}  = (n-2)/(n+2)$. Thus  $w_{\rm re} > 0$ requires higher dimensional operators dominating the minimum. More exotic possibilities for the physics of reheating involve resonant production of particles, tachyonic instabilities,  inhomogeneous modes and turbulence (see \cite{Allahverdi:2010xz} for a review). Recent numerical studies indicate that for all these cases $0 \lesssim w_{\rm re} \lesssim  1/4$ \cite{Podolsky:2005bw}; we will mainly focus on  this range while carrying out our analysis.  We note that instant thermalisation to radiation corresponds to $w_{\rm re} = 1/3$, and this is hard to achieve in practice. For the sake of completeness, we will take a very broad range $-1/3 \leq w_{\rm re} \leq  2/3$ (recall that  $w \leq -1/3$ gives an inflationary epoch) for the analysis in the next section.

 As discussed in the introduction, it is possible to trade the parameter  $N_{\rm re}$ for the last reheat temperature $T_{\rm re}$ in Eq.~\eqref{two}. For a fixed value of $w_{\rm re}$, the   equation  then relates the last reheating temperature to the spectral index.  The relationship  between $N_{\rm re}$ and
$T_{\rm re}$ for K\"ahler moduli inflation can be easily obtained from the analysis in \cite{BLH}. Section  4.2 of \cite{BLH} provides expressions for the energy density at the beginning and end of each epoch of the post-inflationary history of K\"ahler moduli inflation. Using these and incorporating the effect of the reheating epoch we find the Hubble constant at the end of the reheating epoch to be
\begin{equation} 
   H(\hat{t}) = { {M_{\rm pl} W_0^{3}} \over {16 \pi \cv^{9/2} (\ln \cv)^{3/2}} }{ \rm{exp}}\bigg( - { 3 \over 2} (1 + w_{\rm re} ) N_{\rm re} \bigg)~.
\end{equation}
Here the exponential factor takes care of the effects of reheating, and $N_{\rm re} = 0$ corresponds to the instant reheating case.  
Combining this with the usual relationship between the associated energy density and the reheating temperature, $ 3 M_{\rm pl}^{2} H^2 ( \hat{t}) = \rho (\hat{t}) \approx {  \pi^{2} \over 30 } g_{*} T_{\rm re}^4$ (where $g_{*}$ is the effective number of degrees of freedom of the Standard Model sector); and taking $\cv \approx 10^{5}$, $g_{*} \approx 100$ (the exact value of $g_{*}$ has only logarithmic dependence) we find
\begin{equation} 
\label{reheat_eq}
  T_{\rm re} \simeq  10^{3} \ \ \rm{exp} \bigg( - { 3 \over 4} (1 + w_{\rm re} ) N_{\rm re} \bigg) \ \ {\rm{GeV}}~.
\end{equation}
In the next section, we will present our main results by analysing the dependence of $N_{\rm re}$ and $ T_{\rm re}$ on scalar spectral index $n_s$. 

\begin{figure}[!tbp]
  \centering
  \subfloat[]{\includegraphics[width=0.5\textwidth]{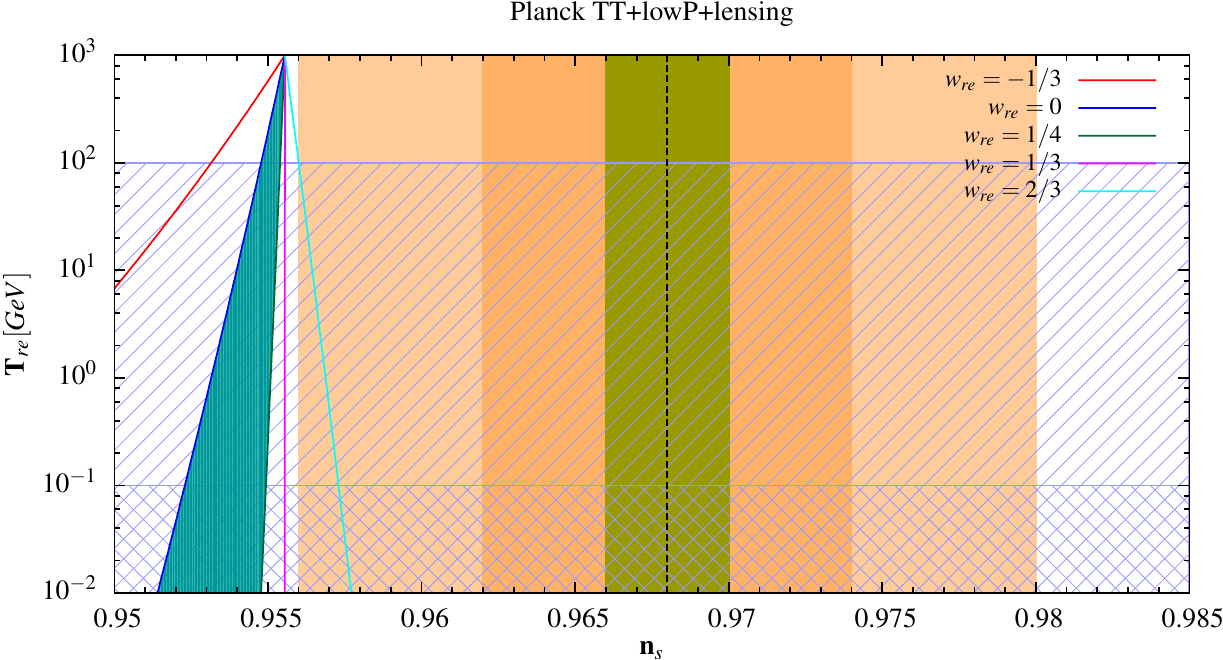}\label{fig:f1}}
  \hfill
  \subfloat[]{\includegraphics[width=0.5\textwidth]{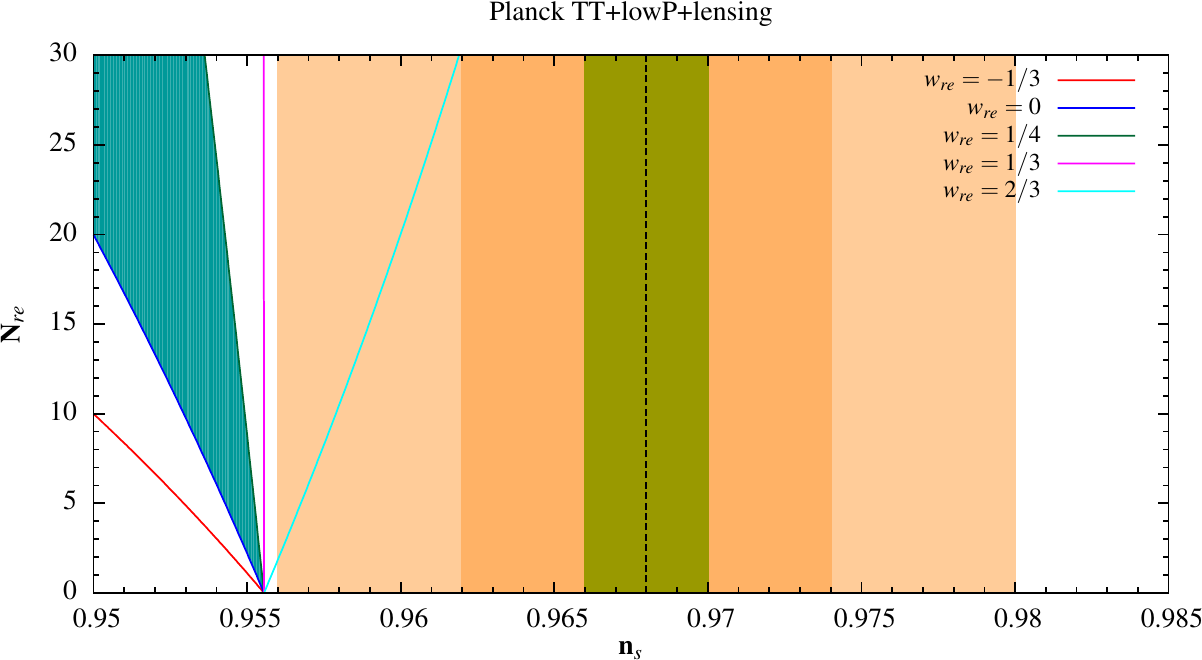}\label{fig:f2}}
  \caption{Plot (a) shows $T_{\rm re}$ as a function of $n_s$ for different values of the equation of state parameter $w_{\rm re} $:  $w_{\rm re}  = -1/3$ red, $w_{\rm re}  = 0$ blue, $w_{\rm re}  = 1/4$ green, $w_{\rm re}  = 1/3$ magneta and  $w_{\rm re} = 2/3$ cyan. These lines meet each other at the point corresponding to $N_{\rm re} = 0$. The vertical dashed black  line shows the PLANCK central value ($n_s = 0.968$) for TT+lowP+lensing data \cite{planck2015}. The  dark brown band corresponds to the $1$-$\sigma$ region $(\Delta n_s \sim 0.006)$ and  the light brown band to the $2$-$\sigma$ region. The  green band marks the projected future  $1$-$\sigma$ sensitivity region with $\Delta n_s \sim 0.002$; assuming that the central value remains unchanged \cite{Abazajian:2016yjj, Finelli:2016cyd}. The blue region corresponds to the parameter space for a physically well motivated reheating scenario with $0 < w_{\rm re} < 1/4$. The horizontally marked mesh region is excluded from BBN constraints; $T_{\rm re} \gtrsim 10$ MeV. On the other hand, the region with right slanted lines requires a non-standard scenario for cosmology at the electroweak scale, $T_{\rm EW} = 100$ GeV. Plot (b) shows $N_{\rm re}$ as a function of $n_s$ with lines and regions marked with the same colour coding as plot (a).}
\end{figure}

\section{Comparison to Observations }
\label{results}

We now have all the ingredients necessary to compare the model predictions with the observational data. For fixed values of the equation of state parameter $w_{\rm re}$, we plot $T_{\rm re}$  and $N_{\rm re}$ as a function of the spectral index $n_s$ (using Eq.~\eqref{two} and Eq.~\eqref{reheat_eq}) in Fig.~\ref{fig:f1} and Fig.~\ref{fig:f2} respectively.  We choose five benchmark values for $w_{\rm re}$ in the range discussed in the previous section  ($-1/3 \leq w_{\rm re} \leq  2/3$). Recall that canonical reheating corresponds to $w_{\rm re}= 0$, and  $w_{\rm re} =1/3$  corresponds to instantaneous thermalisation to radiation. Although the region $w_{\rm re} > 1/3 $ is not very well motivated physically, we also present  plots for $w_{\rm re}= 2/3$ for the purposes of illustration. Numerical simulations of reheating  suggest $0 < w_{\rm re} < 1/4$ \cite{Podolsky:2005bw};  we explicitly mark this range in the plots. 
\begin{figure}[h]
\centering
\includegraphics[width=0.5\textwidth]{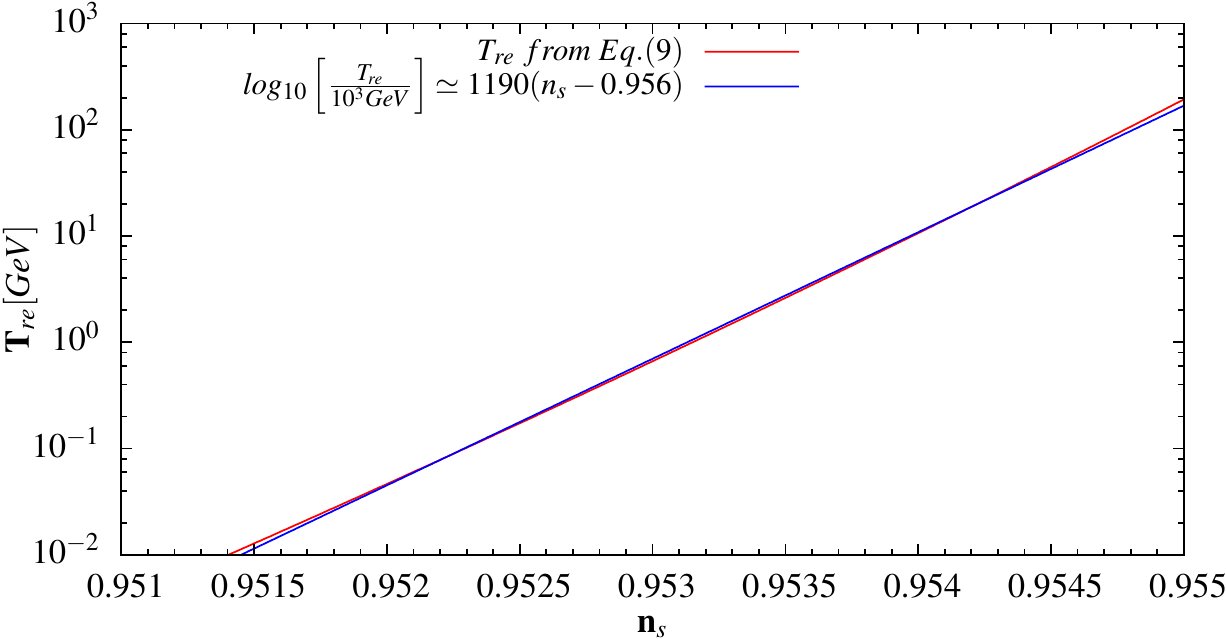}
\caption{Fitting the reheating temperature in terms of the spectral tilt $n_s$ for $w_{\rm re} = 0$ case.}
\label{fig:f5}
\end{figure}

Fig.~\ref{fig:f1} and Fig.~\ref{fig:f2} are based on observational data from PLANCK 2015 (TT+lowP+lensing)  for the $\Lambda$CDM + $r$ model; $n_s = 0.968 \pm 0.006$ at $1$-$\sigma$ \cite{planck2015}. It is clear from the plots that the
 predicted value of $n_s$ is outside the PLANCK $2$-$\sigma$ lower bound for the physically well motivated range of $w_{\rm re}$. It is only for  $w_{\rm re} = 2/3$ that the predicted $n_s$ can become consistent with the $2$-$\sigma$ limit; but this  requires an extended reheating epoch with $N_{\rm re} \gtrsim 30$. If we demand $T_{\rm re} \gtrsim 100$ GeV so as to have a standard scenario for electroweak phase transition, even this comes under a lot of tension. The only way the model can be viable with a realistic reheating scenario is if the observed value of $n_s$ shifts to  become further red tilted in future observations\footnote{We would like to emphasise that the statements being made are for the theoretical predictions being made on the basis of the analysis of \cite{BLH}, global embeddings of the model can potentially change these. }. For  comparison  with various data sets, it is useful to obtain a simple  relationship between $n_s$ and $T_{\rm re}$, and this can be done using least square fitting method. For canonical reheating ($w_{\rm re} = 0$) we find: $\text{log}_{10}(T_{\rm re}/10^3~ \text{GeV}) \simeq 1190 (n_s - 0.956)$ (see Fig.~\ref{fig:f5}). This clearly exhibits the difficultly in matching
 with data since the reheating temperature in the model is bounded by $10^{3}~\text{Gev}$.
 
\begin{figure}[h]
  \centering
  \subfloat[]{\includegraphics[width=0.5\textwidth]{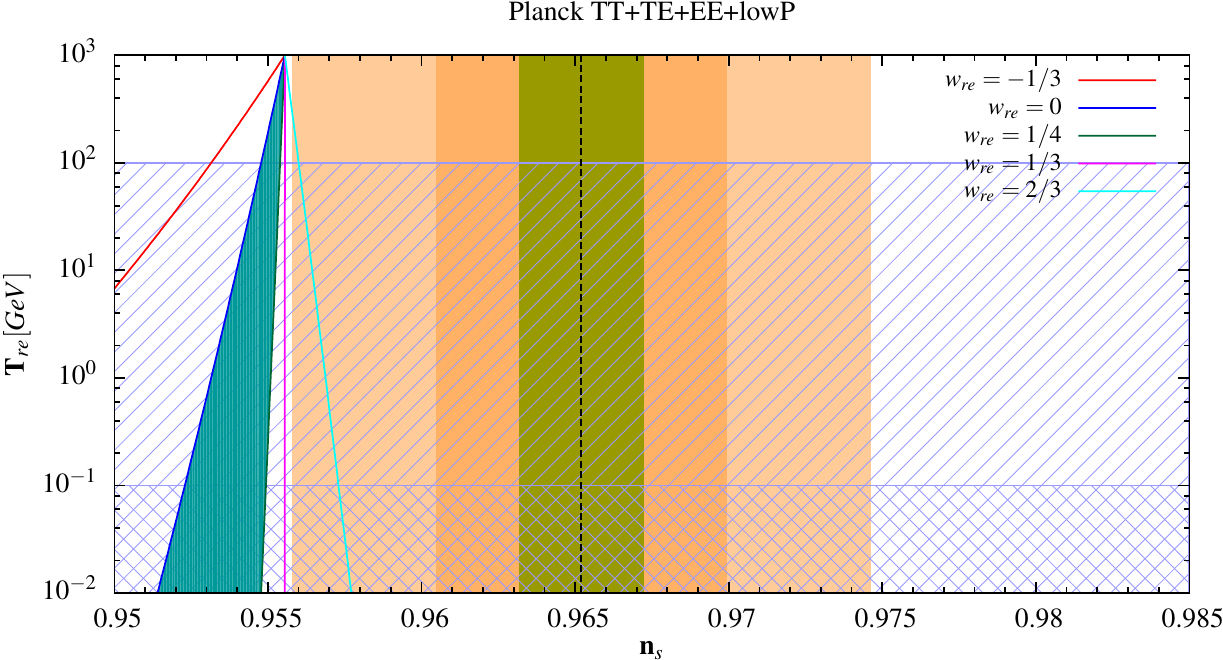}\label{fig:f3}}
  \hfill
  \subfloat[]{\includegraphics[width=0.5\textwidth]{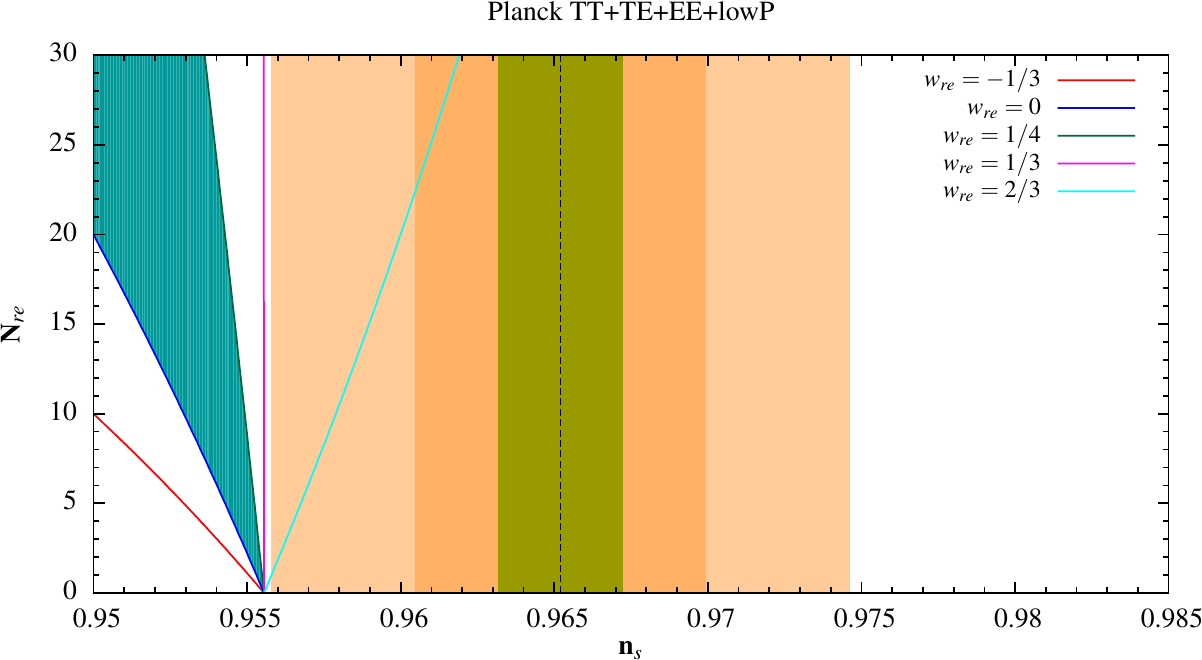}\label{fig:f4}}
  \caption{Plots for Planck (TT,TE,EE+lowP) data with the same colour coding as Fig. 1.  }
\end{figure}

Next, let us consider  Planck (TT,TE,EE+lowP) data for which the central value for $n_s$ becomes smaller $n_s \simeq 0.965$, but the associated error also decreases. Our analysis is summarised in  Fig.~\ref{fig:f3} and Fig.~\ref{fig:f4}. It is easily seen that the conclusion is unchanged. 

\begin{figure}[t]
  \centering
  \subfloat[]{\includegraphics[width=0.5\textwidth]{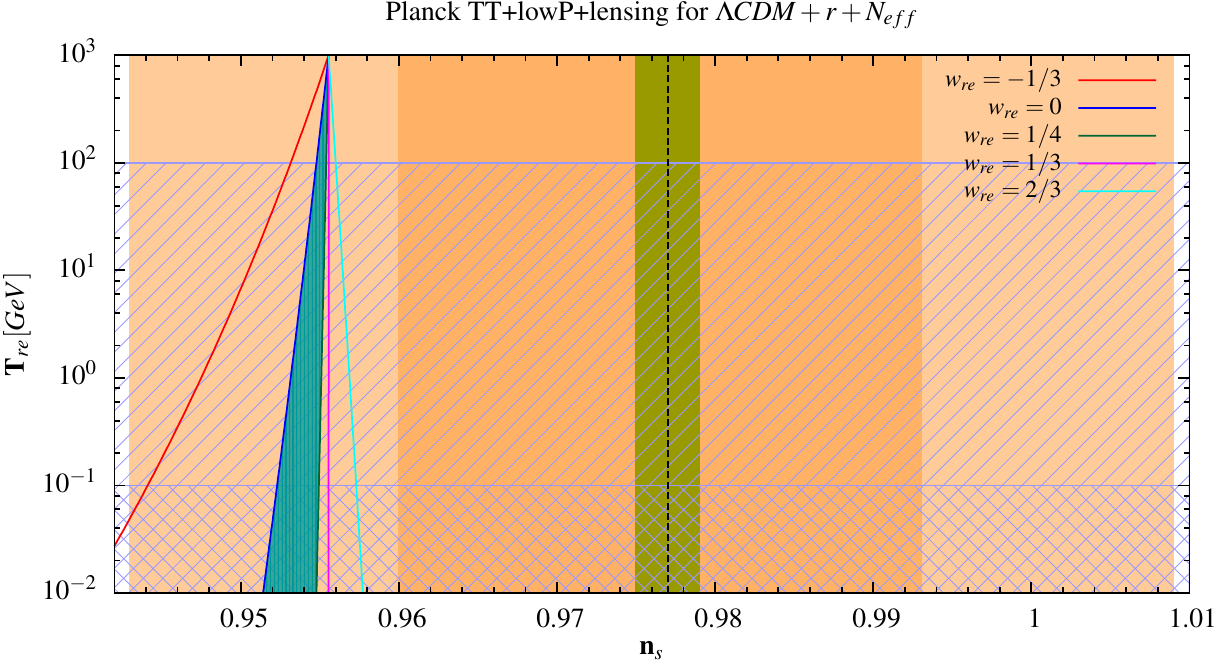}\label{dr1}}
  \hfill
  \subfloat[]{\includegraphics[width=0.5\textwidth]{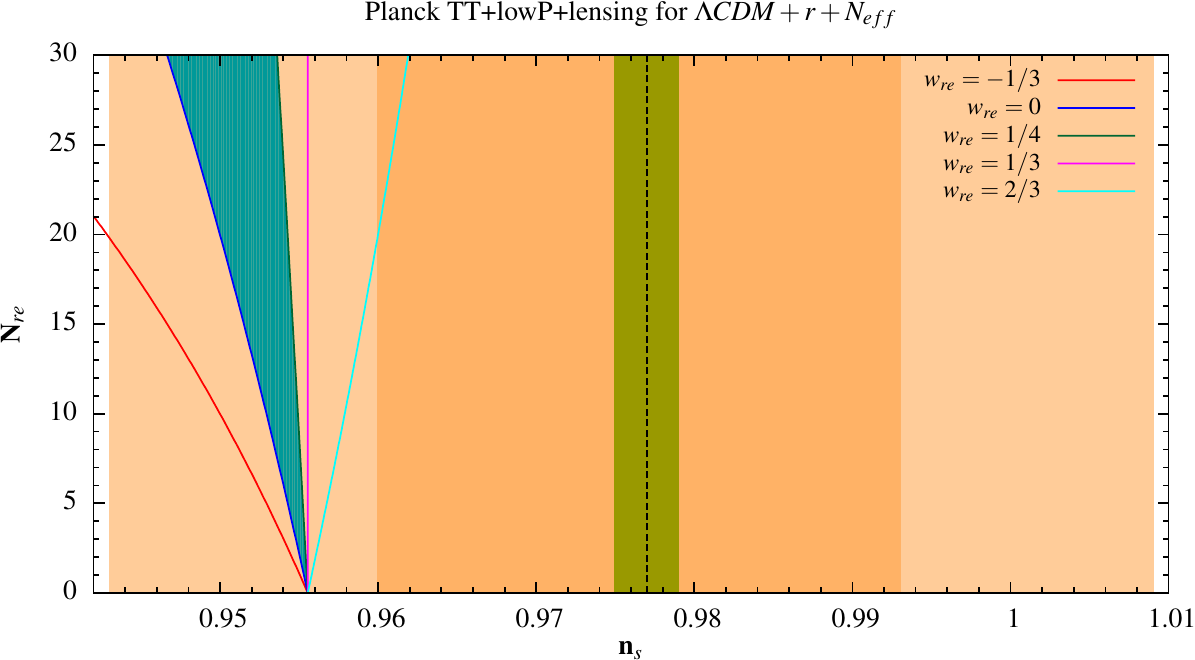}\label{dr2}}
  \caption{Plots for Planck data (TT+lowP+lensing+r+$N_{\rm eff}$) with dark radiation incorporated, the colour coding is same as Fig. 1.  }
\end{figure}

  Dark radiation is a generic feature of string models. In LVS, the axionic partner of the volume modulus is a natural
candidate for dark radiation \cite{micu1}. Comparison to Planck data (TT+lowP+lensing+r+$N_{\rm eff}$) with dark radiation\footnote{In this case, the central value of the marginalised $\Delta N_{\text{eff}} = 0.24$.} included in the post-inflationary history is presented in Fig.~\ref{dr1} and Fig.~\ref{dr2}. Note that with this the predicted value of $n_s$ is  within the 
$2$-$\sigma$ bound for $N_{\rm re} = 0$ (this was previously noted in \cite{micu}), and remains within it for large values of $N_{\rm re}$. But we note that for this data set, the 2-$\sigma$ range is large compared to the other sets.

 Future experiments will bring down $\Delta n_s$, and might as well shift the central value of $n_s$. But with the current  measurements of the spectral index we see that the model can be viable only for an exotic reheating scenario or with dark radiation. We stress that due to the existence of a matter dominated post-inflationary epoch, the predicted value $n_s$ becomes smaller as we measure the cosmologically relevant modes at smaller number of e-folds. The effect of reheating just exacerbates the problem further. If the background cosmological model is extended from $\Lambda$CDM + $r$, the constraint can be relaxed in certain cases, but that is highly dependent on what extra physics is added. 

\section{\Large{Conclusions}}

In this paper, we have incorporated the effect of the post inflationary history (the epoch of domination by moduli particles and reheating)  of  K\"ahler moduli inflation to extract the theoretical prediction for the spectral tilt of the model. We have found that for the model to be consistent with either Planck (TT+lowP+lensing) or Planck (TT,TE,EE+lowP) data, one requires an exotic epoch of reheat $(w_{\rm re} \approx 2/3)$. With dark radiation Planck (TT+lowP+lensing) for $\Lambda CDM+r+N_{\rm eff}$, the model is within the $2$-$\sigma$ range even after the effects of reheating are incorporated. While we have focussed on a single model in this paper, the results exhibit the importance  of carrying out a similar analysis  for any model of inflation while confronting it with precision data. A crucial input for our analysis was the contribution to $N_{*}$ from the epoch in which the energy density of the universe is dominated by cold moduli particles.  To compute this contribution for a model it is necessary to embed the model in a compactification (where the masses and widths of the moduli fields can be determined).  Thus to confront precision data, ``global embedding'' of models (as in \cite{micu}) is absolutely necessary. We note that our analysis is not limited to the case of K\"ahler moduli inflation. The effect is relevant for any inflation model with late decaying scalar field dominating the energy density at the end of inflation (see e.g. \cite{Ellis:2017jcp}). We would like to emphasise that future  experiments like ground based CMB-S4 experiment \cite{Abazajian:2016yjj}, and satellite based experiment CORE \cite{Finelli:2016cyd} are going to measure the spectral index of the CMB  with a projected error of $\Delta n_s\sim 0.002$ $(1$-$\sigma)$; therefore, analysis in the spirit of the present work is going to become more and more important.

\section*{\Large{Acknowledgements}}
KD would like to sincerely thank Kumar Das for discussions and working on a project related to the topic discussed here. Both KD and AM are partially supported by Ramanujan Fellowships funded by SERB, DST, Govt. of India. SB is supported by a fellowship from CSIR, Govt of India.

\end{document}